\documentclass[cits]{PoS}

\usepackage{amsmath, amsthm, amsfonts, amssymb}
\usepackage{graphicx}
\usepackage{multirow}
\usepackage{float}
\usepackage{afterpage}

\newcommand\Poincare {Poincar\'e\ }
\newcommand\Lame {Lam\'e\ }
\newcommand\Backlund {B\"{a}cklund }
\newcommand\Schrodinger {Schr\"{o}dinger }
\newcommand\diag {\mathrm{diag}}
\newcommand\sign {\mathrm{sign}}
\newcommand\arctanh {\mathrm{arctanh}}

\title{Elliptic String Solutions in AdS$_3$ and Elliptic Minimal Surfaces in AdS$_4$}

\ShortTitle{Elliptic String Solutions in AdS$_3$ and Minimal Surfaces in AdS$_4$}

\author{\speaker{Georgios Pastras}%
        \thanks{In memoriam of Prof. Ioannis Bakas}\\
       Department of Physics, School of Applied Mathematics and Physical Sciences, National Technical University, Athens 15780, Greece\\
NCSR ``Demokritos'', Institute of Nuclear and Particle Physics
15310 Aghia Paraskevi, Attiki, Greece\\
       E-mail: \email{pastras@mail.ntua.gr}, \email{pastras@inp.demokritos.gr}}

\abstract{Non-linear sigma models defined on symmetric target spaces have a wide set of applications in modern physics, including the description of string propagation in symmetric spaces, such as AdS or dS, or minimal surfaces in hyperbolic spaces. Although it is difficult to acquire solutions of these models, due to their non-linear nature, it is well known that they are reducible to integrable systems of the family of the sine- or sinh-Gordon equation. In this study, we develop a method to invert Pohlmeyer reduction for elliptic solutions of the reduced system, implementing a relation between NLSM solutions and the eigenstates of the $n = 1$ \Lame problem. This method is applied to produce a family of classical string solutions in AdS$_3$, which includes the spiky strings, as well as hoop string solutions with singular evolution of their angular velocity and radius, which are interesting in the framework of holographic dualities. Furthermore, application of this method produces a wide family of static minimal surfaces in AdS$_4$, which includes helicoids and catenoids, and which are interesting in the framework of the Ryu-Takayanagi conjecture and the understanding of the emergence of gravity as an entropic force related to quantum entanglement statistics. The developed formalism allows the study of the area of the minimal surfaces and geometric phase transitions between them, which are relevant to confinement-deconfinement phase transitions.}

\FullConference{Corfu Summer Institute 2016 "School and Workshops on Elementary Particle Physics and Gravity"\\
                 31 August - 23 September, 2016\\
                 Corfu, Greece}

\begin{document}

\section{Introduction}
\label{sec:Introduction}

The AdS/CFT correspondence is a broad framework that connects gravitational theories in spaces with AdS asymptotics to conformal field theories defined in the boundary of AdS. As a weak/strong duality, AdS/CFT has many applications in the description of strong coupling phenomena in the boundary theory, through simple computations in the weakly coupled gravitational theory. In this context, classical string solutions in AdS spaces have provided insight into the correspondence and the properties of the boundary theory \cite{Gubser:2002tv,Tseytlin:2004xa,Plefka:2005bk}. A more systematic approach, introduced by Alday and Maldacena \cite{Alday:2007hr,Alday:2009yn}, allows the computation of gluon scattering amplitudes from classical string solutions with boundary conditions connected to the gluon momenta.

Similarly, the holographic correspondence can be used for the understanding of the emergence of gravity in the bulk theory from the dynamics of the boundary CFT. The more modern approach to this field manages to explain the similarity of the black hole physics to the laws of thermodynamics describing the gravity as an entropic force originating from quantum entanglement statistics. The original conjecture, made by Ryu and Takayanagi \cite{Ryu:2006bv,Ryu:2006ef}, states that the emergent geometry in the bulk theory is such that it depicts the entanglement entropy of a subsystem of the boundary theory to the area of a co-dimension two minimal surface anchored to the boundary at the entangling surface, i.e. the surface separating the subsystem from its environment,
\begin{equation}
{S_{EE}} = {\frac{1}{4{G_N}}} \, {\rm Area} \left(A^{{\rm{min}}} \right) \, .
\label{eq:RT_conjecture}
\end{equation}
A lot of progress has been made since the original conjecture \cite{Hubeny:2007xt,VanRaamsdonk:2010pw,Takayanagi:2012kg}, including an understanding of the Einstein equations at linear level as equivalent to the first law of entanglement thermodynamics, a trivial statement for any quantum theory \cite{Lashkari:2013koa,Faulkner:2013ica}.

Co-dimension two minimal surfaces in AdS$_4$, as well as classical string solutions in all dimensions are solutions of 
two-dimensional non-linear sigma models defined in a symmetric target space. Solving these models is quite difficult due to their non-linear nature, however, it is well known that they can be reduced to integrable systems. The oldest reduction of this kind is the that of the O$(3)$ NLSM to the sine-Gordon equation \cite{Pohlmeyer:1975nb,Zakharov:1973pp}, known as Pohlmeyer reduction, which has been generalized to O$(N)$ and CP$(N)$ NLSMs \cite{Eichenherr:1979yw,Pohlmeyer:1979ch,Eichenherr:1979uk}. Although the connection between the original and reduced degrees of freedom is non-local, it can be shown that the reduced system can be systematically derived from a local Lagrangian density \cite{Bakas:1993xh,Bakas:1995bm,FernandezPousa:1996hi,Miramontes:2008wt}. Since the emergence of string theory, the Pohlmeyer reduction of NLSMs describing strings propagating in symmetric spaces has been developed \cite{Barbashov:1980kz,DeVega:1992xc,Larsen:1996gn}, including spaces relevant to holographic dualities, such as AdS$_5 \times$S$^5$ \cite{Grigoriev:2007bu,Mikhailov:2007xr,Grigoriev:2008jq} or AdS$_4 \times$CP$^3$ \cite{Rashkov:2008rm}.

The reduced models for NLSM defined on symmetric spaces are typically multi-component generalizations of the sine- or sinh-Gordon equation. The simplest possible systems of interest, which are reducible to a single component system, describe strings propagating in dS$_3$ or AdS$_3$ or static minimal surface in AdS$_4$. The inversion of Pohlmeyer reduction in general is complicated due to its non-local nature and due to the fact that it is not a one-to-one mapping. However, recently, this inversion was achieved for a specific class of solutions of the reduced models via an interesting connection of the NLSM solutions and the band structure of the $n = 1$ \Lame potential \cite{Bakas:2016jxp,Pastras:2016vqu}. In the following, we review this construction and the basic properties of the acquired solutions.

\section{Non-linear Sigma Models and Pohlmeyer Reduction}
\label{sec:Pohlmeyer}

Non-linear sigma models with a single component Pohlmeyer counterpart are those defined on a three dimensional symmetric target space. As an indicative example, in this section, we analyse the case of strings propagating in AdS$_3$. For this purpose, the consideration of AdS$_3$ as a subspace of an enhanced higher-dimensional \emph{flat} space is required. For AdS spaces, this is achieved with the introduction of an extra time-like dimension, implying that AdS$_3$ can be described as a submanifold of $\mathbb{R}^{(2,2)}$, with metric $\eta_{\mu \nu} = \diag\{-1,-1,1,1\}$. Denoting the coordinates in this enhanced space as $Y^{-1}$, $Y^0$, $Y^1$ and $Y^2$, AdS$_3$ is the submanifold
\begin{equation}
Y \cdot Y = - \Lambda^2,
\label{eq:Pohlmeyer_constraint}
\end{equation}
where we use the notation $A \cdot B \equiv \eta_{\mu \nu} {A^\mu }{B^\nu }$. Then, the non-linear sigma model action reads,
\begin{equation}
S = \int {d{\xi_ + }d{\xi_ - } \left( {\partial _ + }Y \cdot {\partial _ - }Y + \lambda \left( {Y \cdot Y + {\Lambda ^2}} \right) \right) } ,
\end{equation}
where $\xi_\pm$ are the left- and right-moving coordinates, ${\xi_ \pm } = \left( {{\xi_1}  \pm {\xi_0} } \right) / 2$.

The equations of motion for the embedding functions $Y$ take the form
\begin{equation}
{\partial _ + }{\partial _ - }Y = \frac{1}{{{\Lambda ^2}}} \left( {{\partial _ + }Y \cdot {\partial _ - }Y} \right)Y ,
\label{eq:eom}
\end{equation}
while a valid solution should also obey the geometric constraint \eqref{eq:Pohlmeyer_constraint}, as well as the Virasoro constraints
\begin{equation}
{\partial _ \pm }Y \cdot {\partial _ \pm }Y = 0 .
\label{eq:Pohlmeyer_Virasoro}
\end{equation}

Pohlmeyer reduction is realized through the introduction of a base of vectors in the enhanced four-dimensional space that includes $Y$, $\partial_+ Y$, $\partial_- Y$ and another space-like unit vector $v_4$, defined to be perpendicular to all other basis vectors. The constraints of the problem imply specific magnitudes and orthogonality for all basis vectors except for the pair $\partial_+ Y$ and $\partial_- Y$. Thus, the only degree of freedom left by the constraints is the angle between the latter, which motivates the definition of the Pohlmeyer field as,
\begin{equation}
{e^\varphi} : = {\partial _ + }Y \cdot {\partial _ - }Y .
\label{eq:Pohlmeyer_a_definition}
\end{equation}
Consistency with the constraints and the equations of motion results in the Pohlmeyer field obeying either the sinh- or cosh-Gordon equation
\begin{equation}
{\partial _ + }{\partial _ - }\varphi = \frac{2}{{{\Lambda ^2}}} { \cosh \varphi } \quad \rm{or} \quad {\partial _ + }{\partial _ - }\varphi = \frac{2}{{{\Lambda ^2}}} { \sinh \varphi } ,
\label{eq:Pohlmeyer_cosh}
\end{equation}
depending on circumstances. The situation is similar for strings propagating in dS$_3$. Pohlmeyer reduction leads to the same equations as in the case of AdS$_3$ with $\Lambda^2 \to - \Lambda^2$.

Few differences appear in the study of static minimal surfaces in AdS$_4$ (i.e. minimal surfaces in H$^3$), too. In this case, the world-sheet coordinates are both space-like, motivating the definition of a complex world-sheet coordinate $z = \left( {{\xi_1} + i {\xi_0} } \right) / 2$, instead of the left- and right-moving coordinates. The extra dimension of the enhanced space has to be time-like and the geometric constraint assumes the same form as in the AdS$_3$ case, given by equation \eqref{eq:Pohlmeyer_constraint}. Pohlmeyer reduction leads solely to the Euclidean cosh-Gordon equation,
\begin{equation}
{\partial }{\bar \partial }\varphi = \frac{2}{{{\Lambda ^2}}} { \cosh \varphi } .
\label{eq:Pohlmeyer_cosh_H}
\end{equation}

The advantage of using a given solution of the Pohlmeyer reduced system to built solutions of the initial NLSM is apparent in the form of the equations of motion \eqref{eq:eom}. Using the definition of the Pohlmeyer field $\varphi$, the latter take the form
\begin{equation}
{\partial _ + }{\partial _ - }Y = \frac{1}{{{\Lambda ^2}}}  {e^\varphi} Y .
\label{eq:eom_varphi}
\end{equation}
Unlike the equations \eqref{eq:eom}, which are non-linear in the embedding coordinates, the system of equations \eqref{eq:eom_varphi} are linear and decoupled.

\section{Elliptic Solutions of the Non-linear Sigma Models}
\label{sec:Solutions}
\subsection{Elliptic Solutions of the Reduced System}
\label{subsec:Elliptic}

The sinh- and cosh-Gordon equations are well-known integrable systems with many known solutions. The usual approach to generate solutions of these systems is the use of \Backlund transformations starting from the vacuum as a seed solution. These techniques though cannot be used in the case of cosh-Gordon equation; Athough it does possesses \Backlund transformations, it does not have a vacuum. Thus, such methods could investigate only a class of classical strings solutions in AdS$_3$ and dS$_3$ and furthermore, they are completely inappropriate for the investigation of minimal surfaces in H$^3$.

In the following, we adopt a different approach that is motivated by the form of the equations of motion \eqref{eq:eom_varphi}. Solutions of the reduced system that depend only on one of the two world-sheet coordinates $\xi_0$ and $\xi_1$ facilitate the solution of the equations of motion via separation of variables. Such solutions are easy to be found, since they solve an one-dimensional version of the corresponding integrable system.

Without loss of generality, we assume that $\varphi$ depends only on the world-sheet coordinate $\xi_1$. In the Lorentzian problems these correspond to static solutions of the corresponding reduced systems. The translationally invariant solutions can be investigated in a similar manner. In the Euclidean problem, there is no discrimination between solutions that depend of $\xi_0$ or $\xi_1$. For this class of solutions, equations \eqref{eq:Pohlmeyer_cosh} or \eqref{eq:Pohlmeyer_cosh_H} reduce to the ordinary differential equation
\begin{equation}
\varphi '' = - s \frac{{{1}}}{\Lambda^2}\left( {{e^{\varphi} } + t{e^{ - {\varphi} }}} \right) ,
\end{equation}
where prime denotes differentiation with respect to $\xi_1$. The parameters $s$ and $t$ were introduced so that all variations of the reduced integrable system can be dealt simultaneously. They take the values $\pm 1$ depending on the case. The latter equation can be integrated to the form
\begin{equation}
\frac{1}{2}{{\varphi '}^2} + s\frac{{{1}}}{\Lambda^2}\left( {{e^{\varphi} } - t{e^{ - {\varphi} }}} \right) = E .
\label{eq:elliptic_energy_conservation}
\end{equation}
Finally, performing the change of variable
\begin{equation}
y :=  - s\frac{{{1}}}{2\Lambda^2}{e^\varphi } + \frac{E}{6},
\label{eq:elliptic_y_definition}
\end{equation}
the equation \eqref{eq:elliptic_energy_conservation} is written as,
\begin{equation}
{{y'}^2} = 4{y^3} - \left( {\frac{{E^2}}{3} + \frac{{{t}}}{\Lambda^4}} \right)y + \frac{E}{3}\left( {\frac{{E^2}}{9} + \frac{{{t}}}{2\Lambda^4}} \right) .
\label{eq:elliptic_p_equation}
\end{equation}

This is the standard form of Weierstrass equation with the moduli taking the specific values
\begin{equation}
g_2 = {\frac{{E^2}}{3} + \frac{{{t}}}{\Lambda^4}} , \quad g_3 = - \frac{E}{3}\left( {\frac{{E^2}}{9} + \frac{{{t}}}{2\Lambda^4}} \right).
\label{eq:elliptic_p_moduli}
\end{equation}
Its general solution in the complex domain is given in terms of the Weierstrass elliptic function $\wp \left( \xi_1 ; g_2 , g_3 \right)$. However, in our case ${e^\varphi }$, and, thus, $y$, must be real, since it is connected to the real embedding functions $Y^\mu$ via the definition of the Pohlmeyer field. Equation \eqref{eq:elliptic_p_equation} may have one or two real solutions in the real domain, depending on the reality of the three roots $e_{1,2,3}$ of the cubic polynomial in the right hand side of \eqref{eq:elliptic_p_equation}. When all three roots are real, let $e_1 > e_2 > e_3$, the equation \eqref{eq:elliptic_p_equation} has two real solutions in the real domain, an unbounded one taking values in $\left[ e_1 , + \infty \right)$ given by
\begin{equation}
y = \wp \left( \xi_1 ; g_2 , g_3 \right)
\label{eq:elliptic_unbounded}
\end{equation}
and a bounded one taking values in $\left[ e_3 , e_2 \right]$ given by
\begin{equation}
y = \wp \left( \xi_1 + \omega_2 ; g_2 , g_3 \right) ,
\label{eq:elliptic_bounded}
\end{equation}
where $\omega_2$ is the imaginary half-period of $\wp \left( \xi_1 ; g_2 , g_3 \right)$. When there is only one real root $e_2$ and two complex ones $e_1$ and $e_3$, equation \eqref{eq:elliptic_p_equation} has only one real solution in the real domain, which is unbounded taking values in $\left[ e_2 , + \infty \right)$ and it is given by expression \eqref{eq:elliptic_unbounded}.

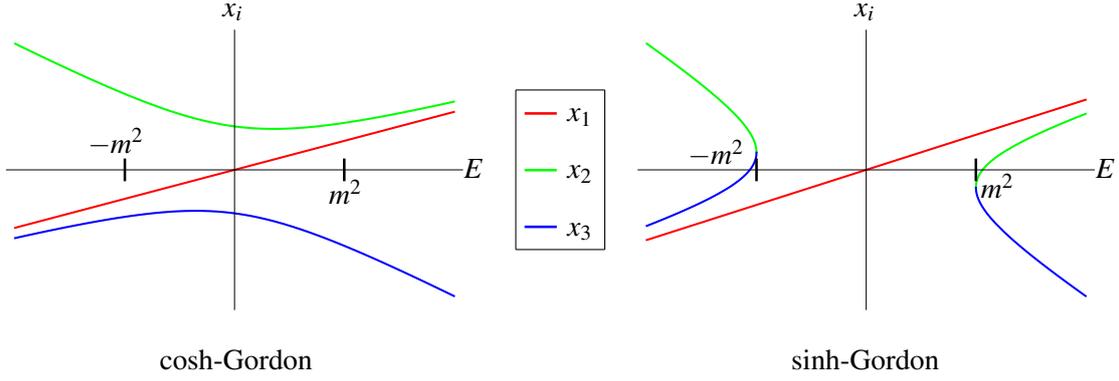
\begin{figure}[ht]
\vspace{10pt}
\begin{center}
\begin{picture}(100,31)
\put(2,5){\includegraphics[width = 0.4\textwidth]{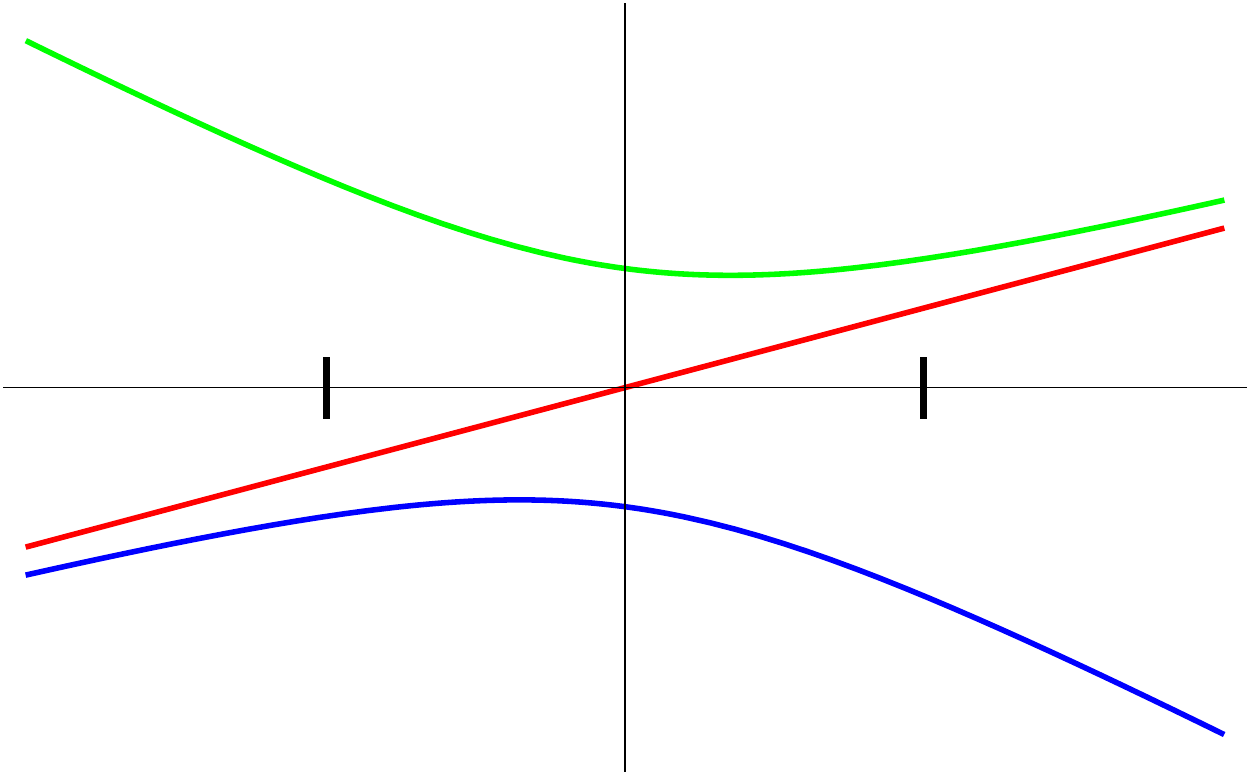}}
\put(57,5){\includegraphics[width = 0.4\textwidth]{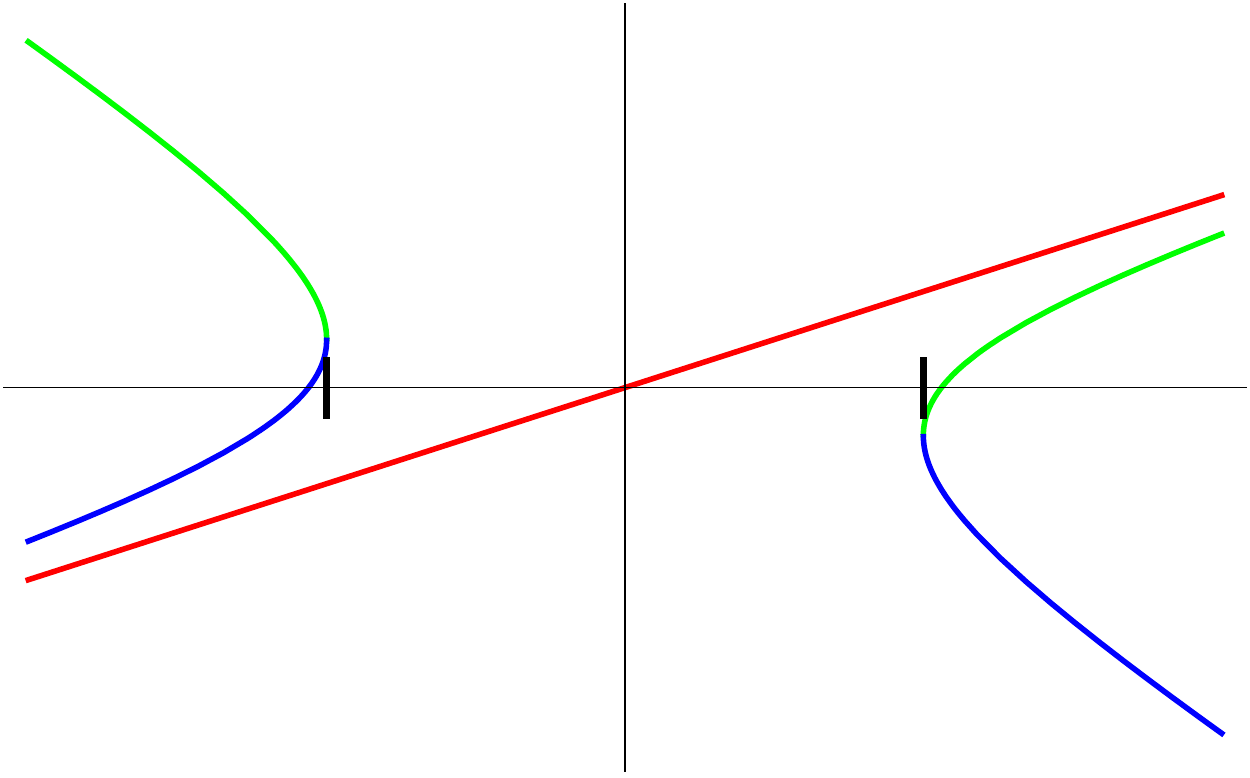}}
\put(46.5,9.5){\includegraphics[height = 0.15\textwidth]{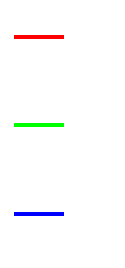}}
\put(9.25,19){$-m^2$}
\put(30.25,14.5){$m^2$}
\put(61.5,18){$-m^2$}
\put(87,15){$m^2$}
\put(51,21.75){$x_1$}
\put(51,16.75){$x_2$}
\put(51,11.75){$x_3$}
\put(21,31){$x_i$}
\put(76,31){$x_i$}
\put(42,16.75){$E$}
\put(97,16.75){$E$}
\put(15.5,0){cosh-Gordon}
\put(70.5,0){sinh-Gordon}
\put(46.5,10.5){\line(0,1){14}}
\put(46.5,10.5){\line(1,0){7.75}}
\put(54.25,10.5){\line(0,1){14}}
\put(46.5,24.5){\line(1,0){7.75}}
\end{picture}
\end{center}
\vspace{-5pt}
\caption{The roots of the cubic polynomial as function of the constant $E$}
\vspace{5pt}
\label{fig:roots}
\end{figure}

The specific values of the moduli $g_2$ and $g_3$ lead to simple expressions for the three roots of the cubic polynomial,
\begin{equation}
{x_1} = \frac{E}{6},\quad {x_{2,3}} =  - \frac{E}{{12}} \pm \frac{1}{4}\sqrt {{E^2} + t{m^4}} .
\label{eq:elliptic_xroots}
\end{equation}
The roots $x_i$ are plotted as functions of the constant $E$ in figure \ref{fig:roots}.
In all cases, a valid solution of the form \eqref{eq:elliptic_unbounded} or \eqref{eq:elliptic_bounded} should be such that the quantity ${e^{\varphi} }$ is not only real, but also positive, so that the world-sheet has the correct signature. This imposes the constraint $- s \left( y - x_1 \right) \geq 0 $, which allows only the bounded \emph{or} unbounded solution, depending on circumstances.

The equation \eqref{eq:elliptic_energy_conservation} can be viewed as the energy conservation for en effective one-dimensional mechanical problem of a point particle under the influence of a hyperbolic potential. In all cases, the solution describes a scattering or oscillatory motion for the point particle, whose ``time of flight'' or period is equal to the real period $2 \omega_1$ of the Weierstrass elliptic function $\wp \left( \xi_1 ; g_2 , g_3 \right)$. Especially in the case of the cosh-Gordon equation, only scattering solutions exist and it turns out that the ``time of flight'' has a global maximum for a specific value of the constant $E = E_0$, satisfying
\begin{equation}
K\left( {k_0 } \right) = 2E\left( {k_0 } \right) , \quad k_0 = \sqrt {\frac{{{e_2 \left(E_0\right)} - {e_3 \left(E_0\right)}}}{{{e_1 \left(E_0\right)} - {e_3 \left(E_0\right)}}}} .
\label{eq:elliptic_E0_analytical}
\end{equation}

\subsection{The Effective \Schrodinger Problems}
\label{subsec:Schrodinger}
All elliptic solutions of equation \eqref{eq:elliptic_p_equation} assume the form
\begin{equation}
- s\frac{{{1}}}{\Lambda^2}{e^\varphi } = 2 \left( \wp \left( \xi_1 + \delta \xi_1 \right) - x_1 \right) ,
\label{eq:Schrodinger_potential}
\end{equation}
implying that the equations of motion are written as
\begin{equation}
\frac{{{d^2 Y^\mu}}}{{d{\xi_1 ^2}}} - \frac{{{d^2 Y^\mu}}}{{d{\xi_0 ^2}}} = 2 \left( \wp \left( \xi_1 + \delta \xi_1 \right) - x_1 \right) {Y^\mu } ,
\end{equation}
where $\delta \xi_1$ vanishes for the unbounded solution while it is equal to $\omega_2$ for the bounded one.

These equations can be solved via separation of variables. We let
\begin{equation}
{Y^\mu }\left( \xi_0 , \xi_1  \right) := {\Sigma ^\mu }\left( \xi_1  \right){{\rm T}^\mu }\left( \xi_0  \right) 
\label{eq:Schrodinger_separation}
\end{equation}
to yield a pair of ordinary differential equations,
\begin{align}
 - \frac{{{d^2}\Sigma^\mu}}{{d{\xi_1 ^2}}} + 2 \left( \wp \left( \xi_1 + \delta \xi_1 \right) - x_1 \right) \Sigma^\mu  &={\kappa^\mu} \Sigma^\mu , \label{eq:Schrodinger_sigma_1} \\
- \frac{{{d^2}{\rm T^\mu}}}{{d{\xi_0 ^2}}} &= {\kappa^\mu}{\rm T^\mu} . \label{eq:Schrodinger_tau_1}
\end{align}
The equations above can be viewed as a pair of effective \Schrodinger problems with identical eigenvalues, one with a trivial flat potential and another one being the $n = 1$ \Lame problem. The solutions to these problems do not have the physical interpretation of a wavefunction, therefore no normalization condition is required for the effective wavefunctions.

Had we used a translationally invariant solution of the reduced system, the situation would be identical with $\Sigma$ and $\rm T$ interchanged. Finally, in the case of minimal surfaces in H$^3$ the situation is also similar, the difference being the fact that the eigenvalues of the pair of effective \Schrodinger problems are opposite instead of equal.

The solutions of the $n = 1$ \Lame problem are well known. They are given by the expressions
\begin{equation}
{y_ \pm } \left( {x ; a} \right) = \frac{{\sigma \left( {x \pm a} \right)}}{{\sigma \left( x \right) \sigma \left( \pm a \right)}}{e^{ - \zeta \left( \pm \alpha  \right)x}} \quad \mathrm{or} \quad {y_ \pm }\left( {x;a} \right) = \frac{{\sigma \left( {x + {\omega _2} \pm a} \right)\sigma \left( {{\omega _2}} \right)}}{{\sigma \left( {x + {\omega _2}} \right)\sigma \left( {{\omega _2} \pm a} \right)}}{e^{ - \zeta \left( { \pm a} \right)x}} ,
\label{eq:Schrodinger_lame_eigenstates}
\end{equation}
depending on whether we study the unbounded or bounded potential. In both cases the corresponding eigenvalues are
\begin{equation}
\lambda = - \wp \left( a \right) .
\label{eq:Schrodinger_lame:eigenvalues}
\end{equation}
In the case of three real roots, the eigenfunctions corresponding to eigenvalues obeying $\lambda < - e_1$ or $ - e_2 < \lambda < - e_3$ are real and exponentially diverging in either plus or minus infinity. The same holds in the case of one real root for $\lambda < - e_2$. The eigenfunctions corresponding to the complementary regions, $\lambda > - e_3$ or $ - e_1 < \lambda < e_2$ and in the case of one real root $\lambda > - e_2$, are complex conjugate to each other and have the form of a Bloch wave. Furthermore, the eigenfunctions obey the following ``normalization'' condition
\begin{equation}
{y_ + }{y_ - } = c \left( \wp \left( x \right) - \wp \left( a \right) \right),
\label{eq:Schrodinger_lame_normalization}
\end{equation}
where $c = 1$ in the case of the unbounded potential and $c = \sign \left( {e_3} - \wp \left( a \right) \right)$ in the case of the bounded potential.

\subsection{Construction of the Solutions}
\label{subsec:construction}

Following section \ref{subsec:Schrodinger}, we have acquired the general solution of the equations of motion \eqref{eq:eom_varphi}. The construction of a classical string solution or a minimal surface requires the appropriate selection of four such solutions of the equation of motion, one for each embedding function in the enhanced space, which also satisfy the geometric constraint \eqref{eq:Pohlmeyer_constraint} and the Virasoro constraints \eqref{eq:Pohlmeyer_Virasoro}.

The simplest possible construction would involve eigenfunctions corresponding to a single eigenvalue of the \Schrodinger problems for all four components. It turns out that such a construction is not possible, and, thus, the simplest possible construction involves two distinct eigenvalues. The geometric constraint and the form of the metric in the enhanced four-dimensional space specify the sign of these two eigenvalues; in the case of strings propagating in AdS$_3$ the eigenvalues have to be of the same sign, whereas in the case of strings propagating in dS$_3$ or minimal surfaces in H$^3$ they have to be of apposite sign.

As an indicative example, we study the case of strings propagating in AdS$_3$ and two positive eigenvalues, but the construction is similar in all cases. The solution falls within an ansatz of the form
\begin{equation}
Y = \left( {\begin{array}{*{20}{c}}
{c_1^ + \Sigma _1^ + \left( {{\xi _1} ; a_1} \right)\cos \left( {\ell _1}{\xi _0} \right) + c_1^ - \Sigma _1^ - \left( {{\xi _1} ; a_1} \right)\sin \left( {\ell _1}{\xi _0}\right)}\\
{c_1^ + \Sigma _1^ + \left( {{\xi _1} ; a_1} \right)\sin \left( {\ell _1}{\xi _0} \right) - c_1^ - \Sigma _1^ - \left( {{\xi _1} ; a_1} \right)\cos \left( {\ell _1}{\xi _0}\right)}\\
{c_2^ + \Sigma _2^ + \left( {{\xi _1} ; a_2} \right)\cos \left( {\ell _2}{\xi _0} \right) + c_2^ - \Sigma _2^ - \left( {{\xi _1} ; a_2} \right)\sin \left( {\ell _2}{\xi _0}\right)}\\
{c_2^ + \Sigma _2^ + \left( {{\xi _1} ; a_2} \right)\sin \left( {\ell _2}{\xi _0} \right) - c_2^ - \Sigma _2^ - \left( {{\xi _1} ; a_2} \right)\cos \left( {\ell _2}{\xi _0}\right)}
\end{array}} \right) .
\label{eq:construction_ansatz}
\end{equation}
The functions $\Sigma _{1,2}^ \pm \left( {{\xi _1} ; a_{1,2}} \right)$ are in general real linear combinations of the functions $y_\pm \left( \xi_1  ; a_{1,2} \right)$ given by \eqref{eq:Schrodinger_lame_eigenstates}. Since the eigenvalues of the $\xi_0$ problem and the $\xi_1$ are identical, it follows that
\begin{equation}
\ell _{1,2}^2 = - \wp \left( {{a_{1,2}}} \right) -2 x_1 .
\label{eq:construction_eigenvalues}
\end{equation}

The geometric constraint \eqref{eq:Pohlmeyer_constraint} assumes the form
\begin{equation}
{\left( {c_1^ + \Sigma _1^ + } \right)^2} + {\left( {c_1^ - \Sigma _1^ - } \right)^2} - {\left( {c_2^ + \Sigma _2^ + } \right)^2} - {\left( {c_2^ - \Sigma _2^ - } \right)^2} = {\Lambda ^2} .
\label{eq:construction_geometric_constraint}
\end{equation}
This equation, combined with the property \eqref{eq:Schrodinger_lame_normalization}, implies that the geometric constraint can be satisfied only if
\begin{align}
c_1^ +  = c_1^ -  \equiv {c_1}, \quad c_2^ +  = c_2^ -  \equiv {c_2} , &\quad c_1^2 = c_2^2 \equiv {c^2} = \pm \frac{\Lambda ^2}{ {\wp \left( {{a_2}} \right) - \wp \left( {{a_1}} \right)} } ,\label{eq:construction_geometric_result_1}\\
\Sigma _{1,2}^ +  = \frac{1}{2}\left( {y_{1,2}^ +  + y_{1,2}^ - } \right),&\quad \Sigma _{1,2}^ -  = \frac{1}{{2i}}\left( {y_{1,2}^ +  - y_{1,2}^ - } \right) , \label{eq:construction_geometric_result_2}
\end{align}
the sign in \eqref{eq:construction_geometric_result_1} depending on the choice of the unbounded or bounded solution of the reduced system. Reality of the solution implies that $y_{1,2}^\pm$ must be complex conjugate to each other, thus, they should correspond to Bloch wave within the allowed bands of the $n = 1$ \Lame potential. This is a general result; \emph{the geometric constraint enforces positive eigenvalues to correspond to Bloch waves and negative eigenvalues to correspond to non-normalizable states within the gaps of the \Lame spectrum}. Furthermore, equation \eqref{eq:construction_geometric_result_1} implies that string solutions corresponding to the unbounded solution of the reduced system obey ${\wp \left( {{a_2}} \right) - \wp \left( {{a_1}} \right)} > 0$, whereas string solutions that correspond to the bounded solution of the reduced system obey ${\wp \left( {{a_2}} \right) - \wp \left( {{a_1}} \right)} < 0$.

It is a matter of simple algebra to show that the Virasoro constraint \eqref{eq:Pohlmeyer_Virasoro} assumes the form
\begin{equation}
\wp \left( {{a_1}} \right) + \wp \left( {{a_2}} \right) =  - x_1 .
\label{eq:construction_Virasoro_result}
\end{equation}

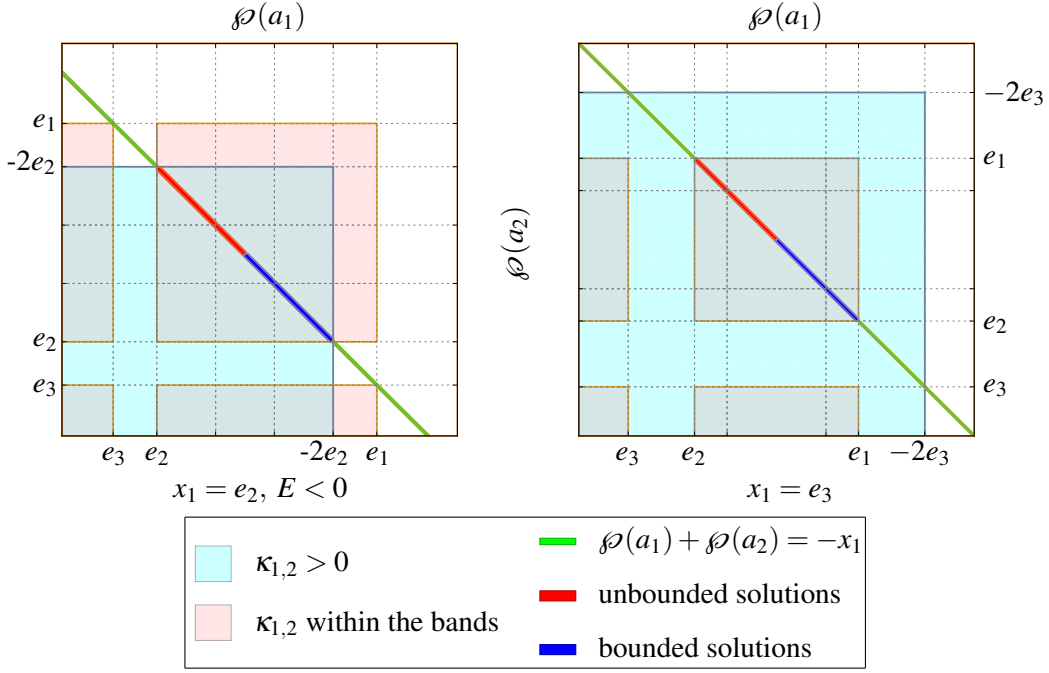
\begin{figure}[ht]
\centering
\begin{picture}(100,60)
\put(10,20){\includegraphics[width = 0.35\textwidth]{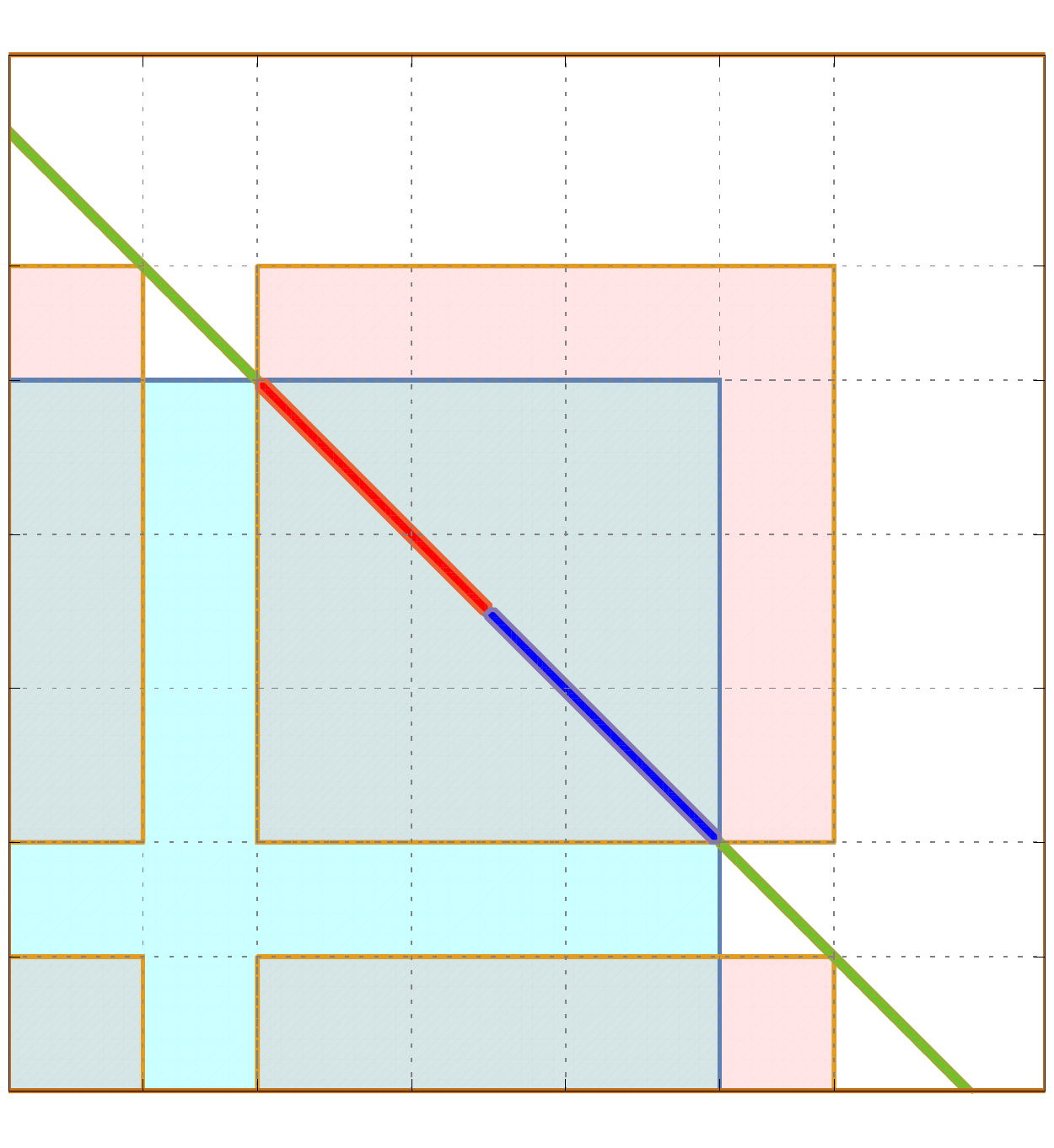}}
\put(55,20){\includegraphics[width = 0.35\textwidth]{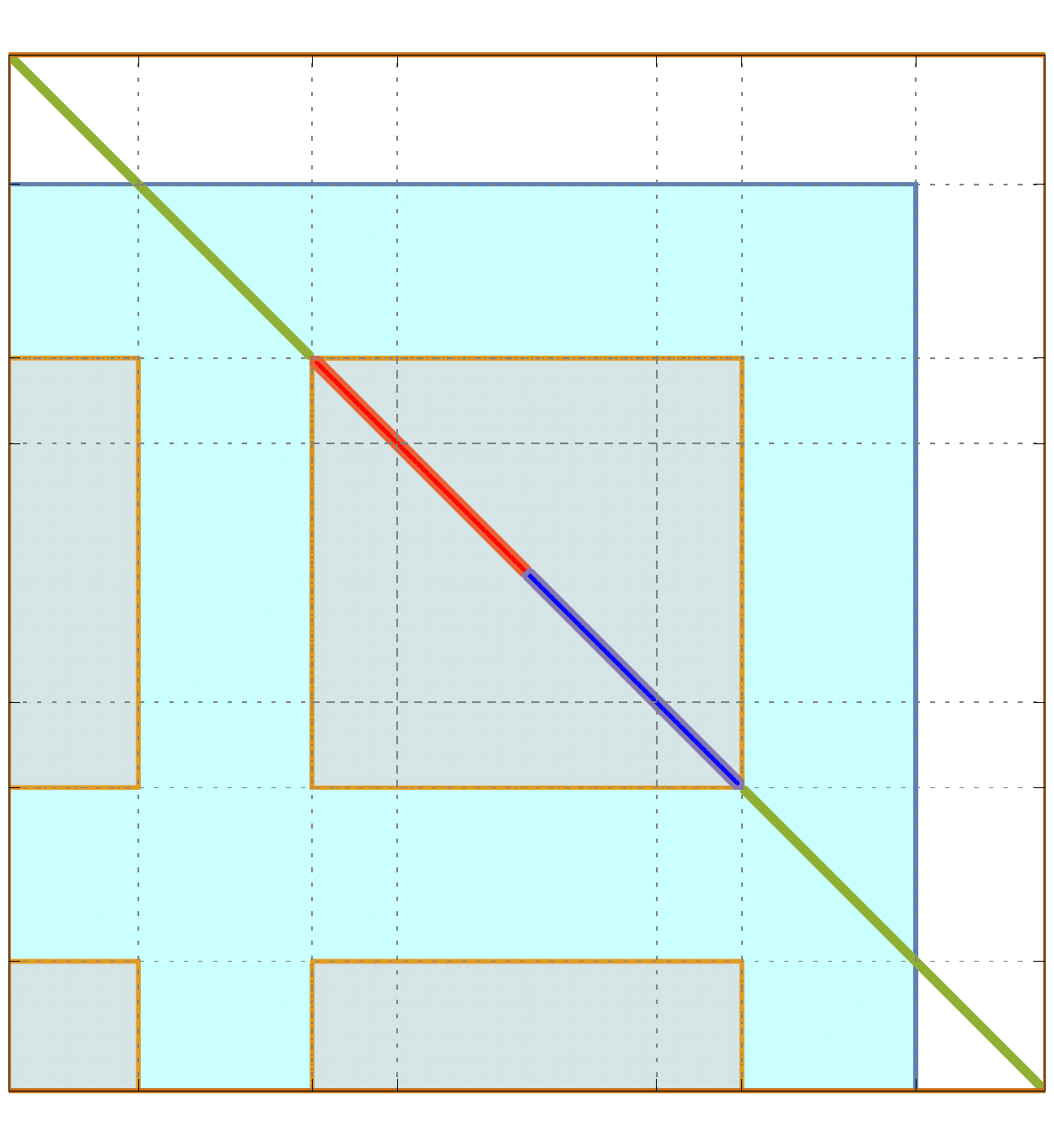}}
\put(21,2){\includegraphics[width = 0.08\textwidth]{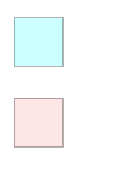}}
\put(51,0){\includegraphics[width = 0.08\textwidth]{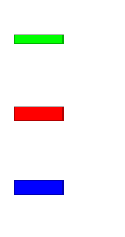}}
\put(7.75,49){$e_1$}
\put(5.5,45.25){-$2e_2$}
\put(7.75,29.75){$e_2$}
\put(7.75,25.75){$e_3$}
\put(90.5,51.25){$- 2 e_3$}
\put(90.5,45.75){$e_1$}
\put(90.5,31.25){$e_2$}
\put(90.5,25.75){$e_3$}
\put(13.75,19.75){$e_3$}
\put(17.5,19.75){$e_2$}
\put(31,19.75){-$2e_2$}
\put(36.75,19.75){$e_1$}
\put(58.5,19.75){$e_3$}
\put(64.25,19.75){$e_2$}
\put(78.75,19.75){$e_1$}
\put(82.25,19.75){$- 2 e_3$}
\put(25,58){$\wp \left( a_1 \right)$}
\put(70,58){$\wp \left( a_1 \right)$}
\put(48.5,35.5){\rotatebox{90}{$\wp \left( a_2 \right)$}}
\put(20,16.5){$x_1 = e_2, \; E < 0$}
\put(70,16.5){$x_1 = e_3$}
\put(27,10){$\kappa_{1,2} > 0$}
\put(27,4.75){$\kappa_{1,2}$ within the bands}
\put(57,12){$\wp \left( a_1 \right)+\wp \left( a_2 \right) = -x_1$}
\put(57,7.25){unbounded solutions}
\put(57,2.5){bounded solutions}
\put(21,1.5){\line(0,1){13.25}}
\put(21,1.5){\line(1,0){59.25}}
\put(80.25,1.5){\line(0,1){13.25}}
\put(21,14.75){\line(1,0){59.25}}
\end{picture}
\vspace{-20pt}
\caption{The pairs of $\wp\left( a_1 \right)$ and $\wp\left( a_2 \right)$ that generate classical string solutions in AdS$_3$ built from eigenstates of the effective \Schrodinger problems corresponding to two distinct positive eigenvalues}
\label{fig:regions_strings_ads}
\end{figure}

Putting everything together, finding a string solution or a minimal surface is equivalent to finding a pair of $\wp \left( {{a_1}} \right)$ and $\wp \left( {{a_2}} \right)$, such that:
\begin{enumerate}
\item They take appropriate values, so that the eigenvalues of corresponding effective \Schrodinger problems \eqref{eq:construction_eigenvalues} have the appropriate sign, 
\item They take appropriate values, so that the corresponding solutions of the $n = 1$ \Lame problem lie within the bands or the gaps of the spectrum, depending on the sign of the corresponding eigenvalues.
\item They obey equation \eqref{eq:construction_Virasoro_result}.
\end{enumerate}
The above constraints can be easily solved graphically, as in figure \ref{fig:regions_strings_ads} for strings in AdS$_3$ and positive eigenvalues. The solutions in general depends on whether the root $x_1$ is identified with $e_1$, $e_2$ or $e_3$.

\section{Properties of the Solutions}
\label{sec:properties}
\subsection{String Solutions in AdS$_3$}
\label{subsec:properties_strings}

To better visualize the form of the constructed solutions, we convert to global coordinates
\begin{equation}
Y^{-1} = \Lambda \sqrt{{1 + {r^2}}} \cos t , \quad Y^0 = \Lambda \sqrt{{1 + {r^2}}} \sin t , \quad Y^1 = \Lambda r \cos \varphi , \quad \quad Y^2 = \Lambda r \sin \varphi .
\end{equation}
The string solutions associated to the unbounded configurations take the parametric form
\begin{equation}
r = \sqrt {\frac{{\wp \left( \xi _1 \right) - \wp \left( {{a_2}} \right)}}{{\wp \left( {{a_2}} \right) - \wp \left( {{a_1}} \right)}}} , \quad t = {\ell _1}{\xi _0} - \arg y_+ \left( \xi_1 ; a_1 \right) , \quad \varphi = {\ell _2}{\xi _0} - \arg y_+ \left( \xi_1 ; a_2 \right) .
\end{equation}
Likewise, for the bounded configurations, the corresponding string solution is
\begin{equation}
r = \sqrt {\frac{{\wp \left( {{a_2}} \right) - \wp \left( {\xi _1 + {\omega _2}} \right)}}{{\wp \left( {{a_1}} \right) - \wp \left( {{a_2}} \right)}}} , \quad t = {\ell _1}{\xi _0} - \arg y_+ \left( \xi_1 ; a_1 \right) , \quad \varphi = {\ell _2}{\xi _0} - \arg y_+ \left( \xi_1 ; a_2 \right) .
\end{equation}

It is clear that in both cases, the solution corresponds to a rigidly rotating string with constant angular velocity $\omega = \ell_2 / \ell_1$. For the unbounded solutions, the angular velocity is smaller than one and the solution extends up to infinite radius, whereas for the bounded ones the angular velocity is greater than one and the solution extends up to maximum radius. This is kinematically expected; for a rigidly rotating configuration in AdS$_3$ with $\omega > 1$, there is a finite value of the radial coordinate $r = \arctanh \left( 1 / \omega \right)$, where the velocity reaches the speed of light. At this radius the solution becomes singular presenting spikes and it turns out that the solution coincides with the AdS$_3$ spiky string solutions \cite{Kruczenski:2004wg}. A bounded solution should obey appropriate periodic conditions so that it is single valued in the enhanced space, namely the angular opening $\delta \varphi$ between two consecutive spikes should equal $\delta \varphi = 2\pi / n$, $n\in \mathbb{Z}$.

For the case $x_1 = e_3$, there is an interesting limit as
$ \wp \left( a_{1,2} \right) \to e_{1,2} $. In this limit, the $n = 1$ \Lame eigenfunctions that appear in the solution correspond to the edge of the allowed bands, and, thus, they become both real and periodic. Therefore, the solution assumes the form $\phi - \omega t = 0$ and it describes a straight string rotating like a rigid rod around its center. This is the Gubser-Klebanov-Polyakov solution \cite{Gubser:2002tv}, which in this context arises as a degenerate limit of a spiky string with two spikes.

In a similar manner, one can construct string solutions corresponding to translationally invariant solutions of the reduced system. Such solutions assume the parametric form,
\begin{equation}
r = \sqrt {\frac{{\wp \left( {{a_2}} \right) - \wp \left( {\xi _0 + {\omega _2}} \right)}}{{\wp \left( {{a_1}} \right) - \wp \left( {{a_2}} \right)}}} , \quad t = {\ell _1}{\xi _1} - \arg y_+ \left( \xi_0 ; a_1 \right) , \quad \varphi = {\ell _2}{\xi _1} - \arg y_+ \left( \xi_0 ; a_2 \right) .
\end{equation}
These solutions can be understood as the outcome of the interchange of the angle $\varphi$ and time applied on a finite spiky string. The solution looks like a hoop string rotating with time-dependent angular velocity and radius. The radius varies between two extrema and the analogue of the spike appears at the maximum radius, where the hoop string reaches the speed of light and gets violently reflected towards smaller radii. The periodic dependence of the enhanced coordinates on the global time coordinate implies that the period of the oscillation should equal $T = 2\pi / n$, $n\in \mathbb{Z}$, similarly to the periodic conditions applied to the finite spiky strings.

Solutions corresponding to negative eigenvalues have quite complicated expressions in global coordinates. If studied in a hyperbolic slicing of AdS$_3$, they look like a periodic spiky structure translating without changing shape. More details are provided in \cite{Bakas:2016jxp}.

\subsection{Minimal Surfaces in H$^3$}
\label{subsec:properties_surfaces}

\begin{figure}[ht]
\centering
\begin{picture}(100,59)
\put(10,19){\includegraphics[width = 0.35\textwidth]{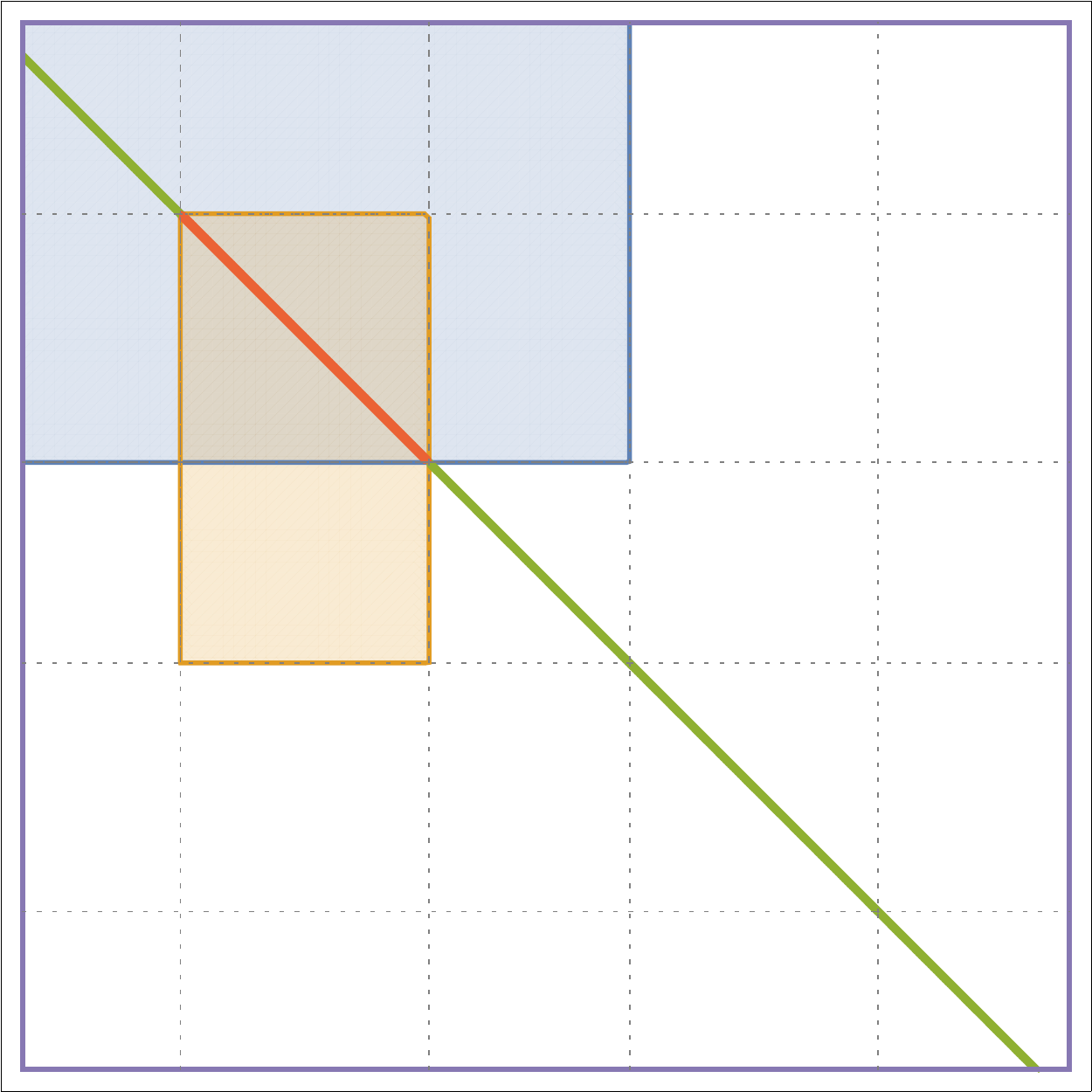}}
\put(55,19){\includegraphics[width = 0.35\textwidth]{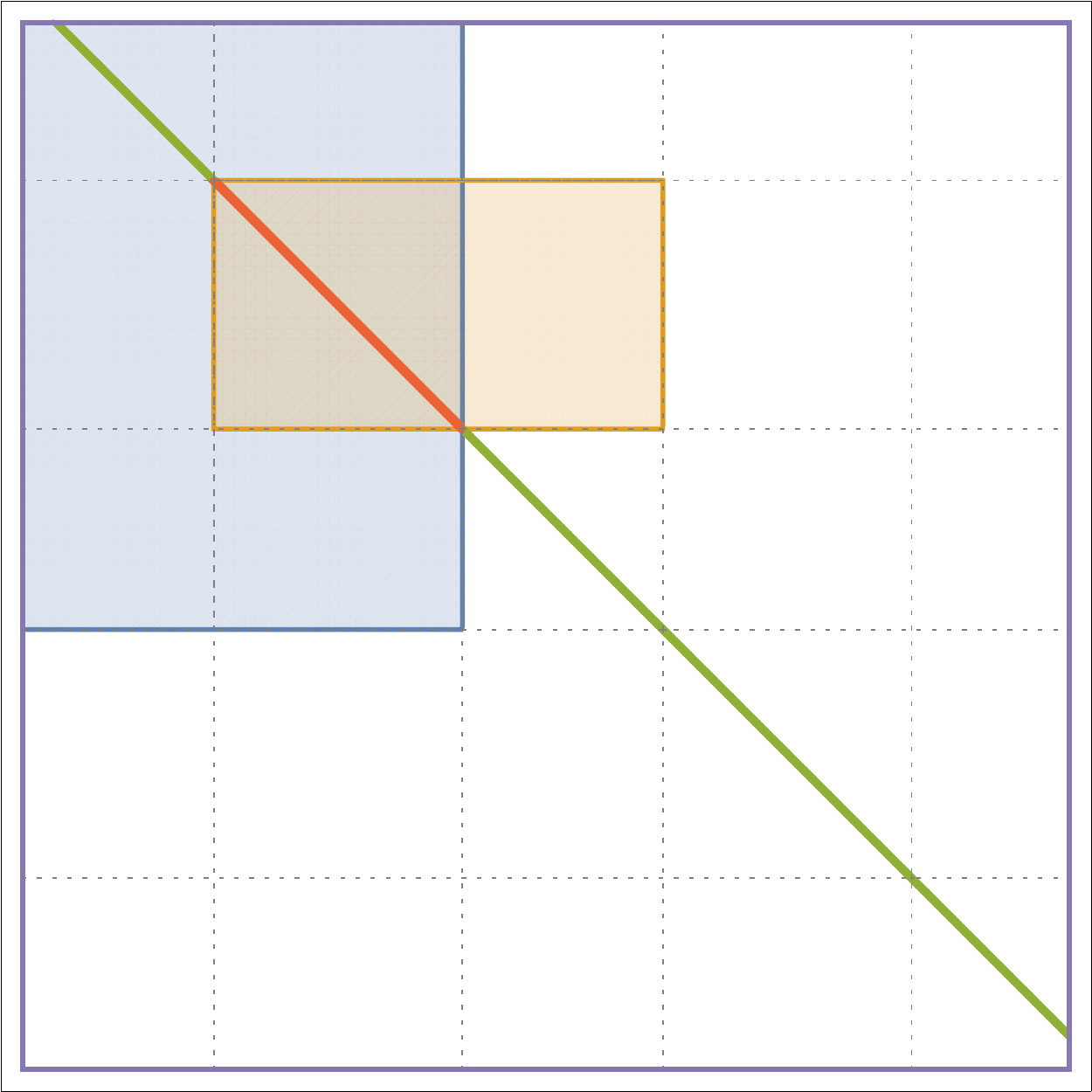}}
\put(19.25,0.25){\includegraphics[width = 0.08\textwidth]{regions_legend_left.pdf}}
\put(49.25,0.5){\includegraphics[width = 0.08\textwidth]{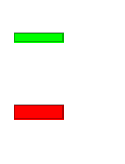}}
\put(7.75,46.75){$e_1$}
\put(5.5,38.75){-$2e_2$}
\put(7.75,32.25){$e_2$}
\put(7.75,24.5){$e_3$}
\put(90.5,47.5){$e_1$}
\put(90.5,39.5){$e_2$}
\put(90.5,33.25){-$2e_2$}
\put(90.5,25.5){$e_3$}
\put(15,16.75){$e_3$}
\put(22.75,16.75){$e_2$}
\put(27.5,16.75){-$2e_2$}
\put(37.25,16.75){$e_1$}
\put(61,16.75){$e_3$}
\put(67.5,16.75){-$2e_2$}
\put(75,16.75){$e_2$}
\put(83.25,16.75){$e_1$}
\put(25,56.25){$\wp \left( a_1 \right)$}
\put(70,56.25){$\wp \left( a_1 \right)$}
\put(48.5,34){\rotatebox{90}{$\wp \left( a_2 \right)$}}
\put(25,13.5){$E < 0$}
\put(70,13.5){$E > 0$}
\put(25.25,8.25){$\kappa_1 > 0$, $\kappa_2 < 0$}
\put(25.25,3){$\kappa_1$ in gap, $\kappa_2$ in band}
\put(55.25,8.25){$\wp \left( a_1 \right)+\wp \left( a_2 \right) = -x_1$}
\put(55.25,3){minimal surface solutions}
\put(19.5,1.5){\line(0,1){9.75}}
\put(19.5,1.5){\line(1,0){62.5}}
\put(82,1.5){\line(0,1){9.75}}
\put(19.5,11.25){\line(1,0){62.5}}
\end{picture}
\vspace{-20pt}
\caption{The pairs of $\wp\left( a_1 \right)$ and $\wp\left( a_2 \right)$ that lead to elliptic minimal surfaces}
\label{fig:regions_surfaces}
\end{figure}

The methods of section \ref{sec:Solutions} can be applied for the construction of static minimal surfaces in AdS$_4$. Two eigenvalues of opposite sign are required, $\kappa_1 < 0$, $\kappa_2 > 0$. The fact that $\kappa_1 < \kappa_2$ enforces them to correspond to the finite gap and band of the $n = 1$ \Lame spectrum, respectively. It turns out that only unbounded solutions can be found. This is expected since bounded surfaces would be shrinkable to a point, and, thus, not minimal. The graphic solution for $\wp \left( a_1 \right)$ and $\wp \left( a_2 \right)$ is depicted in figure \ref{fig:regions_surfaces}. More details are provided in \cite{Pastras:2016vqu}.

The family of elliptic minimal surfaces in H$^3$ is a two-parameter family. One of the parameters is the integration constant $E$; the other is the parameter $\wp \left( a_1 \right)$. Surfaces characterized by the same constant $E$ have identical principal curvatures and they comprise an associate family of minimal surfaces. As a consequence, they share the same local stability properties. The moduli space of the solutions can be visualized as a region bounded by three smooth curves connected in a non-smooth way. These three curves coincide with the boundaries of the segments of solutions in figure \ref{fig:regions_surfaces} and provide interesting limits of the elliptic minimal surfaces:
\begin{itemize}
\item For $\wp\left( a_1 \right) = e_3$, the elliptic minimal surfaces are ruled surfaces, the helicoids in H$^3$.
\item For $\wp\left( a_1 \right) = -2e_2$, the minimal surfaces are surfaces of revolution, the catenoids in H$^3$
\item For $\wp\left( a_1 \right) = e_2$, the elliptic minimal surfaces are conical surfaces.
\end{itemize}

In the following, we use the \Poincare coordinates in order to study some properties of the elliptic minimal surfaces,
\begin{equation}
Y^{-1} = {\frac{1}{{2z}} z^{-1} \left( {{z^2} + {r^2} + {\Lambda ^2}} \right)} , \quad Y^0 = {\frac{1}{{2z}}\left( {{z^2} + {r^2} - {\Lambda ^2}} \right)} , \quad Y^1 = {\frac{\Lambda }{z}r\cos \varphi } , \quad Y^2 = {\frac{\Lambda }{z}r\sin \varphi } .
\end{equation}
The minimal surfaces intersect the boundary when the Weierstrass elliptic function diverges, namely at $\xi_1 = 2 n \omega_1$. Thus, surfaces that are appropriately anchored at the boundary are spanned by $\xi_1 \in \left(2 n \omega_1 , 2 \left( n + 1 \right) \omega_1 \right)$ and $\xi_0 \in \mathbb{R}$, where $n \in \mathbb{Z}$. The trace of the minimal surfaces on the boundary is the union of two logarithmic spirals with the same exponent, namely
\begin{align}
{r} = \Lambda {e^{\omega \varphi }} , &\quad
{r} = \Lambda {e^{\omega \left( {\varphi - \delta \varphi } \right)}} ,\\
\omega ={\ell _1}/{\ell _2}, &\quad \delta \varphi = \pi  - 2\left[ {{\ell _1}{\mathop{\rm Im}\nolimits} {\left( \zeta \left( {{\omega _1}} \right){a_2} - \zeta \left( {{a_2}} \right){\omega _1} \right)} + {\ell _2}{\mathop{\rm Re}\nolimits} {\left( \zeta \left( {{\omega _1}} \right){a_1} - \zeta \left( {{a_1}} \right){\omega _1} \right)}} \right] / {\ell _1}.
\label{eq:boundary_df}
\end{align}
Thus, the entangling curve is completely determined by two parameters. Minimal surfaces that share the same values of $\omega$ and $\delta \varphi$ (or $\delta \varphi$'s that sum to $2\pi$) correspond to identical boundary conditions, and, thus, geometric phase transitions between them are possible. In the following, surfaces with $\delta \varphi > 2 \pi$ are not considered, since they have self-intersections. The special case of the catenoids is an exception to the above, since the entangling curve is the union of two concentric circles. In this case, the boundary conditions are determined by the ratio of the two radii.

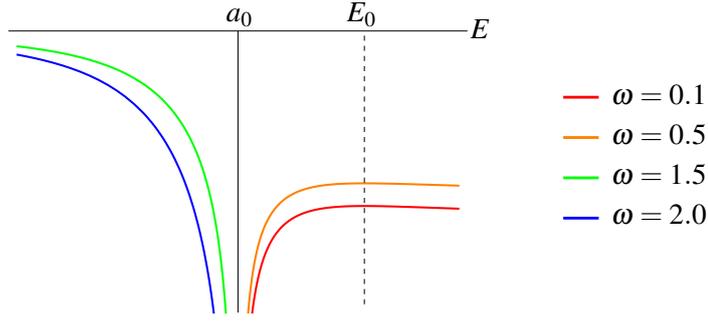
\begin{figure}[ht]
\centering
\begin{picture}(100,28)
\put(22.5,0.5){\includegraphics[width = 0.4\textwidth]{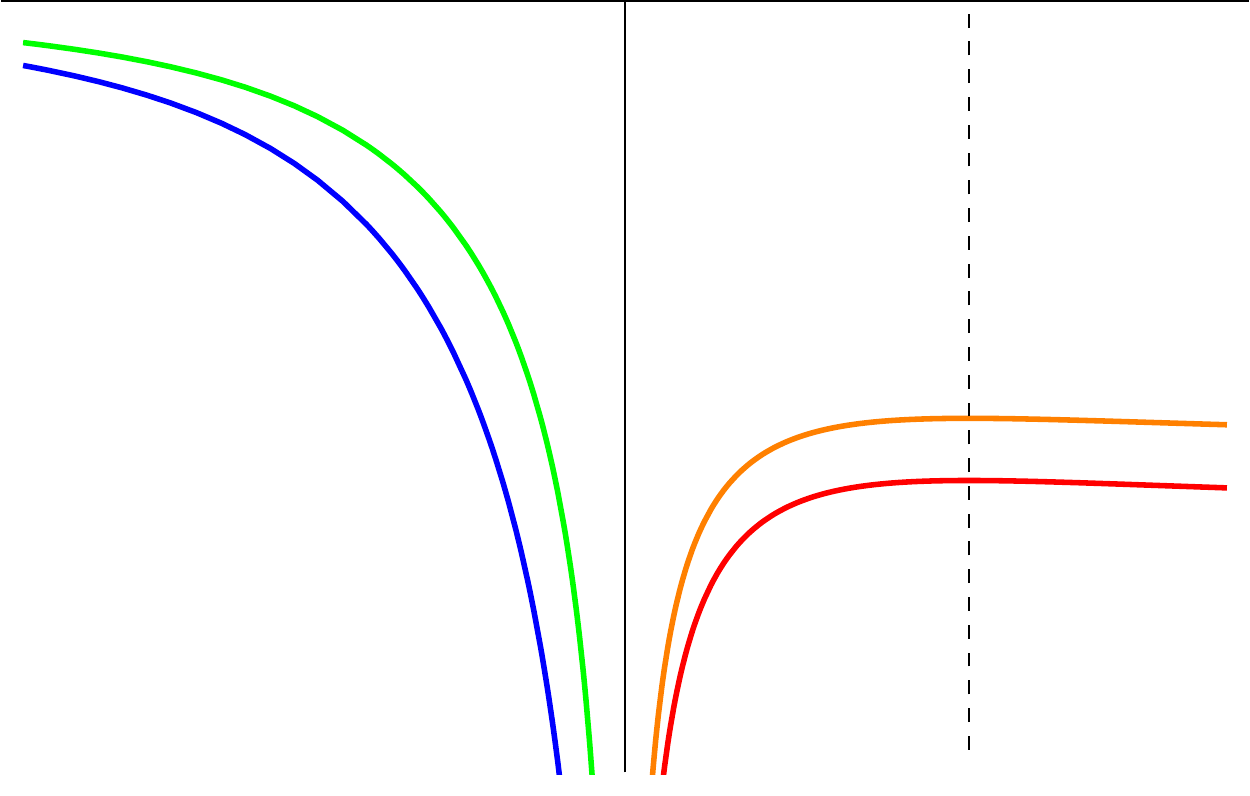}}
\put(70,7){\includegraphics[width = 0.06\textwidth]{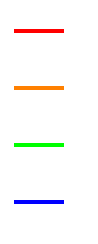}}
\put(62.75,25.25){$E$}
\put(41.5,26.75){$a_0$}
\put(52,26.75){$E_0$}
\put(75,19.5){$\omega = 0.1$}
\put(75,16){$\omega = 0.5$}
\put(75,12.5){$\omega = 1.5$}
\put(75,9){$\omega = 2.0$}
\end{picture}
\vspace{-20pt}
\caption{The coefficient $a_0$ as function of the constant $E$ for various values of the exponent $\omega$}
\label{fig:helicoid_a0}
\end{figure}

The area of the elliptic minimal surfaces, which is connected to the entanglement entropy via the Ryu-Takayanagi conjecture, can be expressed as
\begin{equation}
A = \Lambda L - {\sqrt 2 }{\Lambda ^2} \sqrt {\frac{ {1 - {\omega ^2}} }{E} } \left( {\frac{E}{3} \omega _1} + 2 \zeta \left( \omega_1 \right) \right) \int_{ - \infty }^{ + \infty } {d\varphi}.
\label{eq:properties_area}
\end{equation}
The first term is the usual ``area law''. The second term is the universal constant term, which diverges due to the entangling curve being non-compact. Catenoids are the only exception; in this case the integral $\int_{ - \infty }^{ + \infty } {d\varphi}$ is simply equal to $2 \pi$. We define the quantity
\begin{equation}
a_0 \left( E , \omega \right) := - {\sqrt 2 }{\Lambda ^2} \sqrt {\frac{ {1 - {\omega ^2}} }{E} } \left( {\frac{E}{3} \omega _1 \left(E\right)} + 2 \zeta \left( \omega_1 \left(E\right) \right) \right),
\label{eq:properties_a0}
\end{equation}
which can be used as a measure of comparison for the areas of minimal surfaces corresponding to the same boundary conditions. It can be shown that $a_0$ is negative and has a local maximum at $E = E_0$, where $E_0$ is the value of the constant $E$ that maximizes the real period of the Weierstrass function, given by equation \eqref{eq:elliptic_E0_analytical}. The dependence of $a_0$ on the constant $E$ is depicted in figure \ref{fig:helicoid_a0}.

\begin{figure}[ht]
\centering
\begin{picture}(100,32)
\put(0,0){\includegraphics[width = 0.5\textwidth]{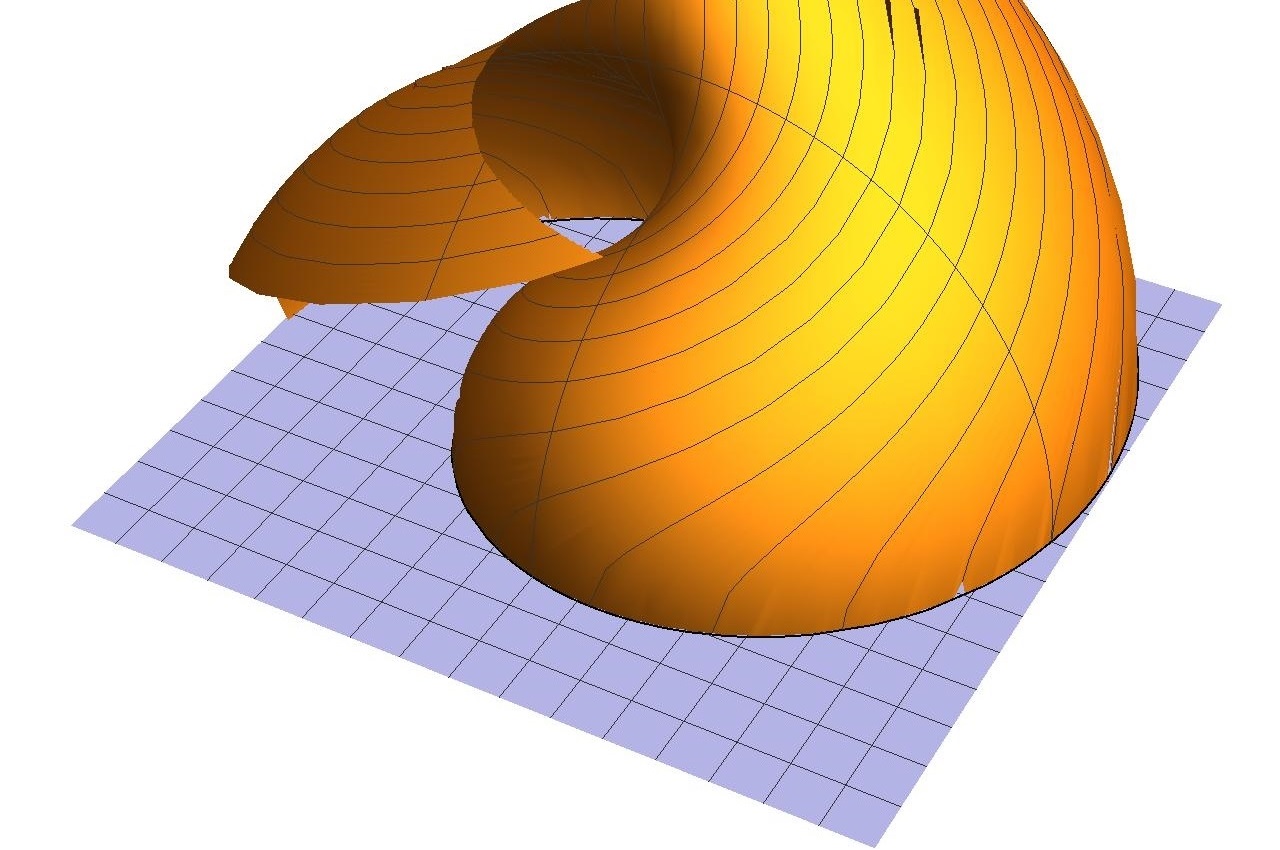}}
\put(50,0){\includegraphics[width = 0.5\textwidth]{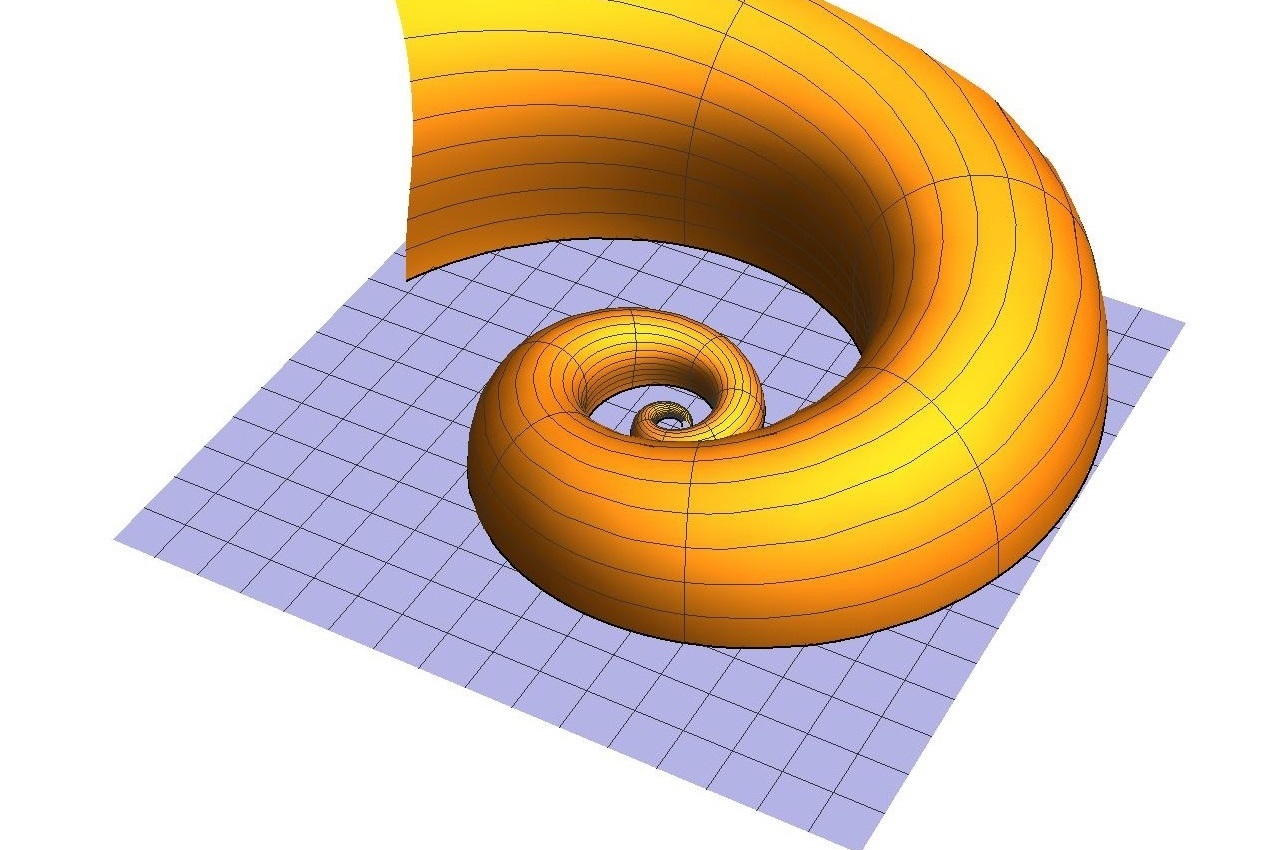}}
\end{picture}
\vspace{-15pt}
\caption{Two elliptic minimal surfaces with the same boundary conditions defined by $\omega = 1 / 4$ and $\delta \varphi = \pi$}
\label{fig:helicoid_PT}
\end{figure}

Moving in the moduli space of elliptic minimal surfaces keeping $\omega$ constant, the constant $E$ varies between $0$ and $1 / \omega - \omega$, the latter corresponding to a helicoid. It turns out that $\delta \varphi$, along such a curve in the moduli space, has the same monotonicity properties as $a_0$. Furthermore, $\delta \varphi$ vanishes for $E = 0$ and is equal to $\pi$ at the helicoid limit. The above imply that there is a critical $\omega_0$, defined as $E_0 = 1 / \omega_0 - \omega_0$; surfaces with $\omega < \omega_0$ do not have a partner surface with the same $\delta \varphi$, whereas surfaces with $\omega > \omega_0$ have two partner minimal surfaces with the same boundary conditions. Two such surfaces are depicted in figure \ref{fig:helicoid_PT}.
Let $E_1 < E_2 < E_3$ be the constants $E$ of the three surfaces with identical boundary conditions. The monotonicity  of $\delta \varphi$ implies that $E_{1,2} < E_0 < E_3$. Equation \eqref{eq:properties_a0}, implies that $a_0 \left( E_1 , \omega \right) < a_0 \left( E_2 , \omega \right) < a_0 \left( E_3 , \omega \right)$. Finally, following \cite{Wang_helicoids}, surfaces with $E > E_0$ are locally unstable. Thus,
\begin{itemize}
\item The surface with $E = E_3$ is locally and globally unstable.
\item The surface with $E = E_2$ is locally stable, but globally unstable.
\item The surface with $E = E_1$ is locally and globally stable.
\end{itemize}
Therefore, as the minimum $E$ surface is always the globally preferred, no geometric phase transitions occur, as the boundary conditions are varied.

\begin{figure}[ht]
\centering
\begin{picture}(100,30)
\put(0,0){\includegraphics[width = 0.5\textwidth]{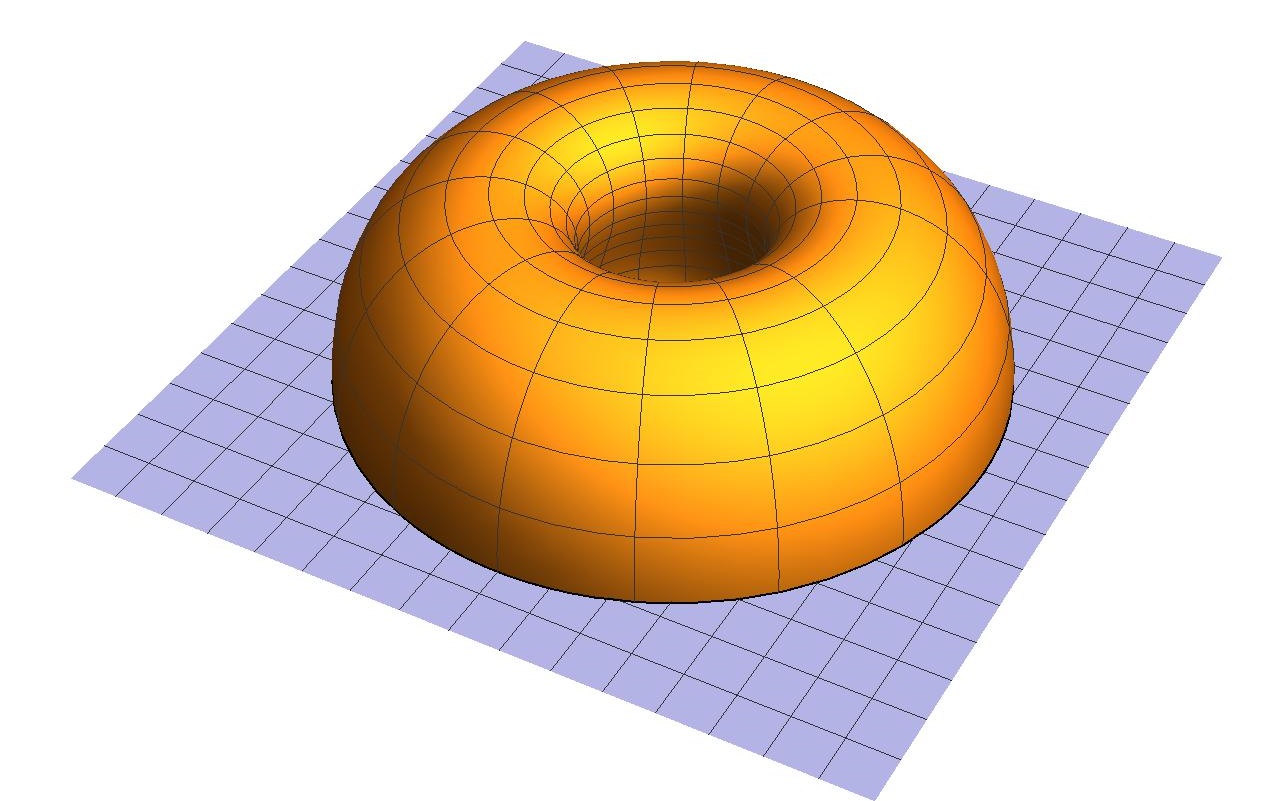}}
\put(50,0){\includegraphics[width = 0.5\textwidth]{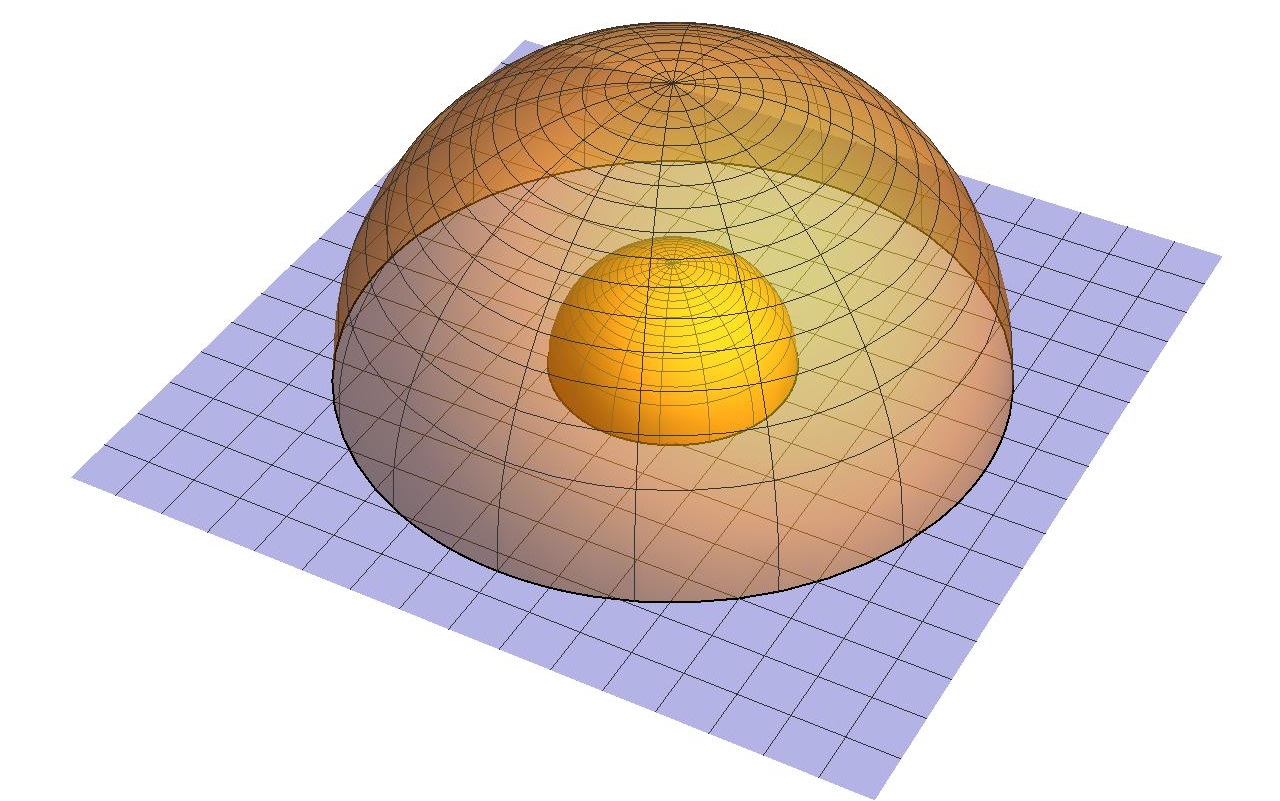}}
\end{picture}
\vspace{-15pt}
\caption{A catenoid and a Goldschmidt minimal surface with the same boundary conditions}
\label{fig:catenoid_PT}
\end{figure}

The analysis for the special case of the catenoids has already been performed in the context of Wilson loops (i.e. \cite{Zarembo:1999bu}). For completeness, we summarize the results in our language. Let $f<1$ be the ratio of the radii of the two circles comprising the entangling curve. Then, for any ratio $f > f_0$, where $f_0$ corresponds to the catenoid with $E = E_0$, there are two catenoids, one locally stable and one locally unstable; for $f < f_0$ there is none. For any value of $f$, there is a Goldschmidt solution, being the union of the minimal surfaces corresponding to each of the boundary circles. Such surfaces are depicted in figure \ref{fig:catenoid_PT}.
Equation \eqref{eq:properties_a0} implies that there is a critical value of the ratio $f_c > f_0$, corresponding to an integration constant $E_c$ satisfying ${E_c}{\omega _1}\left( {{E_c}} \right) + 6\zeta \left( {{\omega _1}\left( {{E_c}} \right)} \right) = 3\sqrt {2{E_c}} $, so that for $f < f_c$ the Goldschmidt minimal surface is the globally preferred one, whereas for $f > f_c$ the stable catenoid is the globally preferred surface. Therefore, there is a geometric phase transition between a catenoid and a Goldschmidt solution.

\section{Discussion}
\label{sec:discussion}

We developed a method, based on the inversion of Pohlmeyer reduction, to construct solutions to NLSMs defined on a three-dimensional symmetric target space, using as a starting point, a specific class of solutions of the reduced theory, namely solutions that depend on only one of the two world-sheet coordinates. For this class of solutions, the equations of motion take the form of four pairs of effective \Schrodinger problems. Each pair consists of a flat potential and and $n = 1$ \Lame potential with connected eigenvalues. Consistency with the constraints select only Bloch waves with positive eigenvalues and non-normalizable states with negative eigenvalues.

Application of the above method to the NLSM describing strings propagating in AdS$_3$ results in a wide class of classical string solutions. This includes the rigidly rotating spiky strings \cite{Kruczenski:2004wg}, rigidly rotating strings that extend to the AdS boundary, as well as hoop strings with periodically varying radius and angular velocity.

It would be interesting to apply this techniques to other target space geometries such as the sphere or the projective plane, as well as to higher dimensional target geometries, where Pohlmeyer reduction results in multi-component generalizations of the sine- and sinh-Gordon equations. Such generalizations would be particularly interesting in the case of space-times that are relevant to holographic dualities, such as AdS$_5 \times$S$^5$ or AdS$_4 \times$CP$^3$.

Inverting Pohlmeyer reduction in the NLSM describing minimal surfaces in H$^3$ results in the construction of a two-parameter family of minimal surfaces, interpolating between the helicoids, catenoids and cusps. These surfaces intersect the AdS boundary at the union of two logarithmic spirals or two concentric circles in the case of catenoids. The local stability properties of this surfaces interestingly are connected with the specific energy that maximizes the ``time of flight'' of a point particle moving under the influence of a sinh potential. Furthermore, there are in general more than one elliptic minimal surfaces that correspond to the same boundary conditions. However, in the more general case of spiral boundary conditions, a geometric phase transition never occurs, unlike the special case of catenoids, where a geometric phase transition between a catenoid and a disjoint surface occurs. Such phase transitions may enlighten the role of entanglement entropy as an order parameter in confinement-deconfinement phase transitions.

An interesting application of the elliptic minimal surfaces in the framework of the Ryu-Takayanagi conjecture would be the verification of the equivalence of the linearized Einstein equations to the first law of entanglement thermodynamics. This has been shown in the case of minimal surfaces corresponding to a circular entangling curve \cite{Lashkari:2013koa,Faulkner:2013ica}. However, these minimal surfaces are special; they have both principal curvature vanishing. Verification of the above equivalence in minimal surfaces with non-trivial curvature, such as the elliptic surfaces presented here, would greatly support the idea of gravity emerging as a quantum entropic force.

\end{document}